\documentclass[preprint,12pt]{elsarticle}

\biboptions{numbers,sort&compress}

\usepackage{graphicx}
\usepackage{amsmath,amssymb}
\usepackage{url}
\usepackage{tikz}
\usepackage{float}
\usetikzlibrary{arrows.meta, positioning, shapes.geometric}
\usepackage[hidelinks]{hyperref}
\usepackage{pdflscape}

\begin{document}

\begin{frontmatter}

\title{A Low-Cost ATmega32-Based Embedded System for Automated Patient Queue and Health Data Management in Private Medical Chambers}


\author[aff1]{Kawshik Kumar Paul\corref{cor1}}
\cortext[cor1]{Corresponding author.}
\ead{kawshikbuet17@gmail.com}

\author[aff1]{Mahdi Hasnat Siyam}
\ead{mahdibuet3@gmail.com}

\author[aff1]{Khandokar Md. Rahat Hossain}
\ead{rahat2975134@gmail.com}

\address[aff1]{Department of Computer Science and Engineering, Bangladesh University of Engineering and Technology (BUET), Dhaka, Bangladesh}

\begin{abstract}
This paper presents a low-cost, stand-alone embedded system that automates patient queue handling and basic health data acquisition for small private medical chambers. The proposed design separates interaction into two physically distinct modules: a patient's self-service corner for entering basic details and measuring vital signs, and a doctor's corner for reviewing the current patient's information and advancing the queue. A single ATmega32 microcontroller coordinates both modules, interfacing with an LM35 temperature sensor, an XD-58C pulse sensor, matrix keypads for data entry, and dual 16$\times$2 LCDs for guided interaction and clinician-side display. Unlike IoT-first approaches that require continuous connectivity and higher deployment overhead, the system operates offline and provides deterministic local operation suitable for resource-constrained settings. Experimental validation shows temperature readings within $\pm 1^{\circ}$C (LM35 range tested), resting pulse readings within $\pm 3$~BPM, and button-to-display latency below 1.2~s, demonstrating reliable real-time performance under limited hardware resources.

\end{abstract}


\end{frontmatter}

\section{Introduction}
Patient management in small private medical chambers is commonly handled through manual tokens, paper registers, or informal verbal ordering, which can become error-prone during peak hours and difficult to audit. In parallel, basic vital measurements (for example temperature and pulse rate) are often taken separately, either by an assistant or by the doctor, which increases consultation overhead and reduces throughput. These constraints motivate a compact, low-cost system that enables guided self-registration, automatic queue assignment, and basic pre-screening of vitals before the patient meets the doctor. 

Many existing solutions for healthcare monitoring emphasize IoT connectivity, mobile applications, and centralized dashboards \cite{irjet2020,rath2021}. While such approaches are effective at scale, they introduce additional cost, network dependency, and operational complexity that may not be appropriate for compact chambers where staff and infrastructure are limited. Similarly, prior embedded queue management systems often target general service counters (e.g., banks, customer care) or rely on additional infrastructure such as PCs or GSM-based workflows \cite{eqc2022,qgsm2013}. A practical solution for small chambers must therefore be inexpensive, offline-capable, easy to operate for non-technical users, and responsive enough to support deterministic real-time interaction. 

Motivated by these requirements, this work proposes a microcontroller-based embedded system that combines (i) a patient-side guided interface for entering basic demographic information and capturing vital signs, and (ii) a doctor-side interface for viewing the current patient's record and advancing the queue with a single control action. The design follows a dual-corner architecture in which the patient booth and the doctor's supervision area remain logically and physically distinct, while still coordinated by a single controller for simplicity and cost reduction. The system is designed to operate deterministically without reliance on external computation or network services and can be extended in future work to optional serial or wireless links if required. The hardware and firmware design choices are guided by microcontroller constraints such as limited SRAM/CPU, ADC resolution, and GPIO availability \cite{microchip2020}, and the use of low-cost analog temperature sensing \cite{ti2022}. 

The main contributions of this work are: \begin{itemize} \item A full embedded design that integrates queue management, guided patient self-registration, and vital sign acquisition on an ATmega32 microcontroller. \item A dual-corner architecture separating patient interaction (input and sensing) from doctor supervision (display and dequeue control) to improve usability in compact clinical settings. \item An offline, stand-alone workflow with deterministic response characteristics suitable for environments where connectivity and infrastructure are limited. \end{itemize}

\section{Related Work}
The integration of microcontroller platforms and network-enabled health monitoring has been widely explored in recent years, particularly through IoT-based architectures that support remote access, centralized dashboards, and continuous parameter logging. Ahmed et al.~\cite{irjet2020} present an IoT-based patient health monitoring approach that reflects this direction by combining edge sensing with network-enabled reporting. Rath et al.~\cite{rath2021} similarly investigate IoT-based monitoring in the context of hospital management, emphasizing system-level monitoring and integration for larger healthcare environments.

In parallel, embedded queue management and patient calling systems have been investigated to reduce manual overhead and streamline service delivery. A representative offline approach is the electronic queue control system in~\cite{eqc2022}, where a microcontroller drives local queue display and control for service counters. Other designs incorporate external communication and infrastructure to coordinate calling and notification; for example, GSM-assisted queue handling for clinical OPD workflows is discussed in~\cite{qgsm2013}. Additional queue management implementations for service environments, including selection-oriented and display-driven embedded systems, are reported in~\cite{arpn2019}. While these approaches demonstrate clear automation benefits, they may introduce deployment complexity (additional infrastructure, connectivity assumptions, or operational dependencies), and they typically do not integrate vital-sign acquisition into the queue workflow. 

Low-cost stand-alone physiological measurement on resource-constrained microcontrollers has also been explored. Systems integrating temperature sensing and pulse measurement with local LCD-based feedback have been reported, for example in~\cite{hrtemp2019}. Such implementations validate that basic vital acquisition can be achieved with inexpensive sensors and simple embedded interfaces. However, many of these designs treat sensing and display as isolated functions, rather than embedding them into a complete clinic workflow that includes guided self-registration, record formation, and queue progression between patient and doctor. 

For resource-constrained embedded deployments, the choice of platform and sensors significantly affects system feasibility and reliability. The ATmega32 microcontroller~\cite{microchip2020} remains a widely used low-cost controller due to its integrated ADC, adequate GPIO availability, and peripheral interfaces that support sensor acquisition and human--machine interfacing. For temperature measurement, the LM35 sensor~\cite{ti2022} is frequently adopted because of its linear voltage-to-temperature characteristic and straightforward analog interface, enabling efficient conversion using the microcontroller ADC. 

In summary, existing work provides strong foundations in (i) IoT-centered health monitoring~\cite{irjet2020,rath2021}, (ii) embedded queue/token systems with varying infrastructure assumptions~\cite{eqc2022,qgsm2013,arpn2019}, and (iii) low-cost embedded vital measurement prototypes~\cite{hrtemp2019}. In contrast, the proposed system focuses on integrating these elements into a single, deterministic, offline embedded workflow for private medical chambers: guided patient-side data entry and vital acquisition, combined with doctor-side queue progression and display, implemented under realistic microcontroller resource constraints~\cite{microchip2020}.

\section{System Architecture}
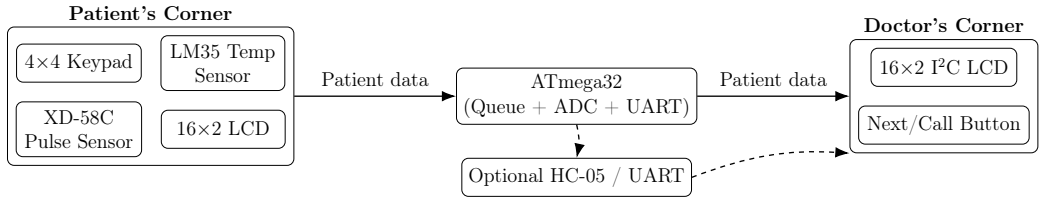
\begin{figure*}[ht]
\centering
\resizebox{\textwidth}{!}{%
\begin{tikzpicture}[
    node distance=1cm and 1.3cm,
    box/.style={rectangle, draw, rounded corners, minimum width=2.6cm, minimum height=0.8cm, align=center},
    group/.style={rectangle, draw, rounded corners, inner sep=5pt},
    arrow/.style={-{Latex[length=3mm]}, thick}
]

\node[group, label=above:{\textbf{Patient's Corner}}] (patientgroup) {
    \begin{tikzpicture}[node distance=0.4cm and 0.4cm]
        \node[box] (keypad) {4$\times$4 Keypad};
        \node[box, right=of keypad] (temp) {LM35 Temp \\Sensor};
        \node[box, below=of keypad] (pulse) {XD-58C\\Pulse Sensor};
        \node[box, right=of pulse] (plcd) {16$\times$2 LCD};
    \end{tikzpicture}
};

\node[box, right=3.4cm of patientgroup] (mcu) {ATmega32\\(Queue + ADC + UART)};

\node[group, right=3.2cm of mcu, label=above:{\textbf{Doctor's Corner}}] (doctorgroup) {
    \begin{tikzpicture}[node distance=0.4cm]
        \node[box] (dlcd) {16$\times$2 I\textsuperscript{2}C LCD};
        \node[box, below=of dlcd] (btn) {Next/Call Button};
    \end{tikzpicture}
};

\draw[arrow] (patientgroup.east) -- node[above, yshift=2pt]{Patient data} (mcu.west);
\draw[arrow] (mcu.east) -- node[above, yshift=2pt]{Patient data} (doctorgroup.west);

\node[box, below=0.7cm of mcu] (bt) {Optional HC-05 / UART};
\draw[arrow, dashed] (mcu.south) to[bend left=12] (bt.north);
\draw[arrow, dashed] (bt.east) to[bend left=12] (doctorgroup.south west);

\end{tikzpicture}%
}
\caption{Compact block diagram showing separation between the Patient's Corner and the Doctor's Corner with ATmega32 at the center.}
\label{fig:system-architecture-block-diagram}
\end{figure*}

The suggested system is divided into two primary modules, the patient's corner and the doctor's corner, as shown in Fig.~\ref{fig:system-architecture-block-diagram}. The separation is intentional: the patient-side workflow (data entry and sensing) and the doctor-side workflow (review and queue control) remain logically and physically distinct, while a single ATmega32 core coordinates the overall operation to keep the design low-cost and easy to deploy in compact medical chambers.

\subsection{Patients' Corner}
The patient interface is implemented as a small self-service booth equipped with simple, low-cost input and sensing peripherals. The patient's corner includes:
\begin{itemize}
    \item two 4$\times$4 matrix keypads for entering patient name (encoded), age, and mobile number;
    \item an LM35 temperature sensor for body temperature acquisition;
    \item an XD-58C pulse sensor for heart rate measurement;
    \item a 16$\times$2 character LCD to guide the patient through step-by-step prompts and display the assigned serial number.
\end{itemize}

The workflow is designed to be deterministic and easy to operate by non-technical users. The LCD prompts the patient through sequential fields (identifier information followed by sensor measurements). After all required fields are entered and stable sensor readings are acquired, the microcontroller assigns a unique serial number and enqueues the complete patient record into an in-memory queue structure for subsequent retrieval by the doctor's corner. This guided flow reduces ambiguity, enforces field completion, and minimizes the chance of partial or invalid records.

\subsection{Doctor's Corner}
On the doctor's table, the supervision interface consists of:
\begin{itemize}
    \item a 16$\times$2 I\textsuperscript{2}C LCD to display the current patient's details (demographics and measured vitals);
    \item a push button to call the next patient (i.e., dequeue).
\end{itemize}

Each press of the doctor's button dequeues the next available patient record and updates the doctor-side display. The I\textsuperscript{2}C LCD reduces the required wiring and GPIO usage compared to a fully parallel LCD interface, which is important given the simultaneous need to interface keypads and analog sensors. The architecture also keeps the queue control action intentionally minimal so that the doctor can advance the queue quickly without navigating complex menus.

\begin{figure*}[!t]
    \centering
    \includegraphics[width=\textwidth]{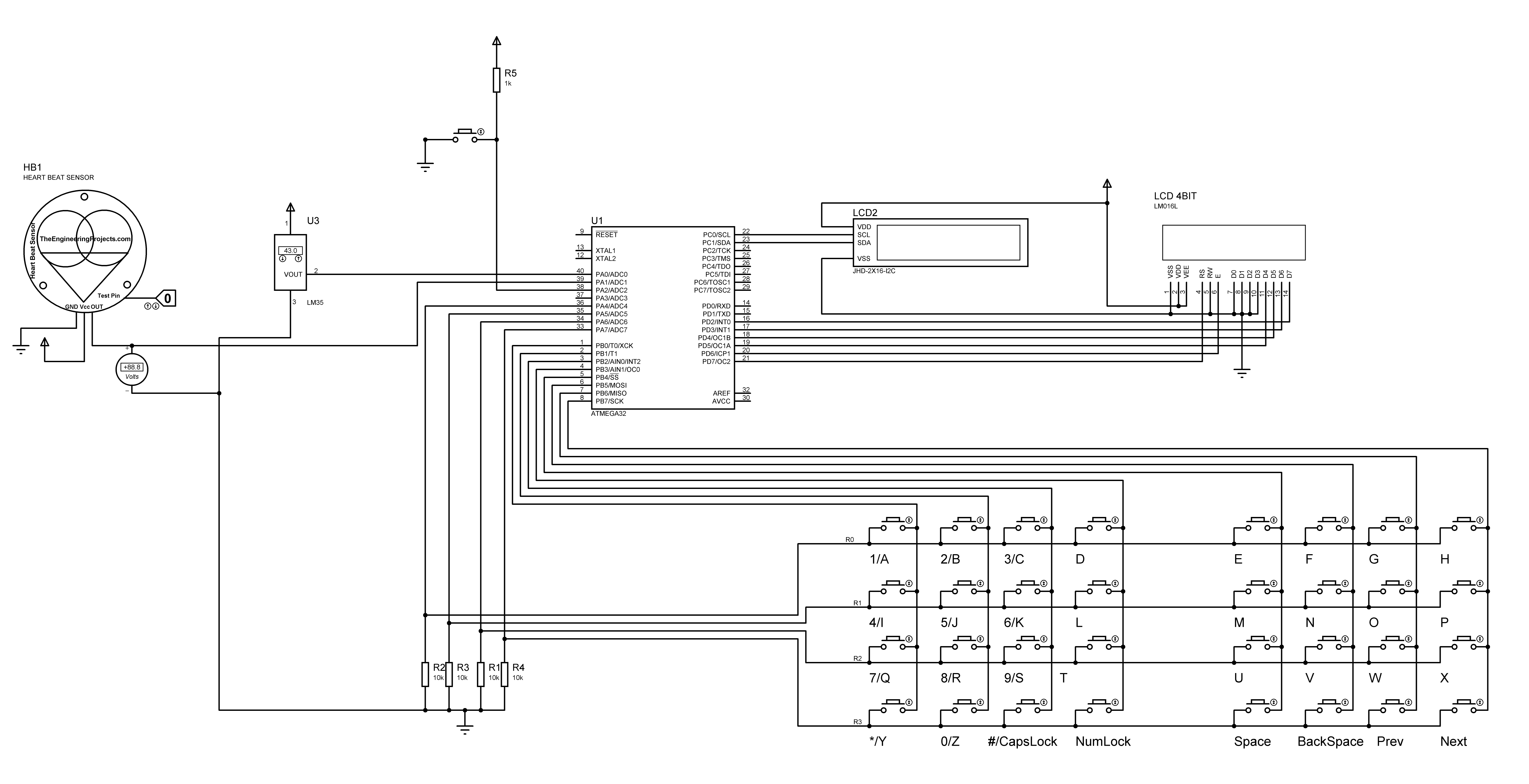}
    \caption{Proteus schematic of the complete hardware architecture including ATmega32, LM35, pulse sensor, LCD, and custom alphanumeric keypad.}
    \label{fig:proteus-complete}
\end{figure*}

Figure~\ref{fig:proteus-complete} shows the complete Proteus schematic corresponding to the architecture. The diagram provides a hardware-level mapping of the two-LCD arrangement (patient-side parallel LCD and doctor-side I\textsuperscript{2}C LCD), keypad scanning lines, ADC sensor connections, and the doctor-side dequeue button. This schematic-level view makes the shared I/O constraints explicit and supports end-to-end verification of peripheral interaction and signal routing before physical deployment. In addition, as illustrated in Fig.~\ref{fig:system-architecture-block-diagram}, the UART interface can be reserved for optional future extensions (for example an HC-05 module) without changing the core offline workflow.

\section{Hardware Implementation}
\subsection{Microcontroller}
The ATmega32 8-bit AVR microcontroller was selected because it provides:
\begin{itemize}
    \item built-in ADC channels for analog sensors (LM35);
    \item UART interface reserved for optional future serial/wireless extensions;
    \item sufficient I/O pins for keypad and LCD control;
    \item support for internal 8 MHz oscillator, avoiding external crystals.
\end{itemize}
In order to lower UART baud-rate error, the internal oscillator was reconfigured during development from 1 MHz to 8 MHz via fuse-bit modification.

\subsection{Sensors}
Firmware can easily convert the 10 mV/$^{\circ}$C output of the LM35A temperature sensor. Periodically, the XD-58C optical pulse sensor is read; heartbeats are identified by thresholding the sampled waveform, and the inter-beat interval is used to calculate beats per minute (BPM).

\subsection{Input/Output Peripherals}
Textual and numerical inputs are collected using two 4$\times$4 matrix keypads. The patient side uses a standard 16$\times$2 LCD (parallel mode) for prompts, and the doctor side uses an I\textsuperscript{2}C 16$\times$2 LCD to save I/O pins. The wiring of these peripherals to the ATmega32 is depicted in the Proteus schematic in Fig.~\ref{fig:proteus-complete}.


\section{Software and Firmware Implementation}
\subsection{Queue Management}
A simple circular buffer was implemented in C to hold patient records. Each record contains:
\begin{itemize}
    \item patient ID / serial number;
    \item basic demographics (age, encoded name or short code, mobile number);
    \item temperature reading;
    \item pulse/BPM reading.
\end{itemize}
Following successful data entry and sensor acquisition, the enqueue operation is initiated. The doctor's button initiates the dequeue process.

\subsection{Sensor Acquisition}
An ADC channel is linked to the LM35. The MCU uses the sensor scale to convert the ADC value to degrees Celsius after periodically sampling the sensor and averaging samples to reduce noise. BPM is calculated as follows: 
\[
\text{BPM} = \frac{60 \times 1000}{T_{\text{beat}}}
\]
where $T_{\text{beat}}$ is the time in milliseconds between two consecutive detected peaks. The pulse sensor is sampled at predetermined intervals; peaks above a threshold are considered beats.

\subsection{LCD Interface}
Step-by-step instructions (``Enter Age', ``Place Finger on Sensor', etc.) are displayed on the patient's LCD, followed by the assigned serial. The dequeued patient's data is displayed on the doctor's LCD.

\subsection{Programming and Configuration} USBasp was used to upload the firmware, which was written in embedded C (AVR-GCC/AVR Studio). To enable more precise UART, fuse bits were changed to choose the internal 8~MHz oscillator.

\section{System Calibration and Debugging}
\subsection{Baud-Rate Error Mitigation}
At 1~MHz, the UART baud generator produced an error that was too high for reliable serial communication at higher baud rates (e.g., 38400~bps). Reliability was significantly increased by using the internal 8~MHz oscillator and double-speed UART.

\subsection{Sensor Validation} LM35 readings were within $\pm 1^{\circ}$C in the 25–40$^{\circ}$C range when compared to a commercial digital thermometer. When compared to a fingertip oximeter, the resting subjects' pulse readings were within $\pm 3$~BPM.

\section{Experimental Results and Discussion}
\begin{table}[!t]
\centering
\caption{System Performance Summary}
\label{tab:experimental_results_table}
\begin{tabular}{l l}
\hline
\textbf{Metric} & \textbf{Observation} \\
\hline
Temperature accuracy & $\pm 1^{\circ}$C (LM35 range tested) \\
Pulse accuracy & $\pm 3$ BPM (resting) \\
Button-to-display latency & $< 1.2$ s \\
Operation mode & Offline, stand-alone \\
\hline
\end{tabular}
\end{table}

The system demonstrates that basic vital acquisition and patient self-registration can be integrated with queue assignment in an offline embedded workflow. Table~\ref{tab:experimental_results_table} summarizes the key observed performance outcomes during validation, emphasizing that the design maintains reliable real-time behavior under limited computational and hardware resources.

For temperature validation, LM35 readings were compared against a calibrated digital thermometer over the tested operating range. Observations remained within $\pm 1^{\circ}$C, which is adequate for non-diagnostic monitoring and pre-screening use-cases in compact settings. For heart rate validation, pulse readings obtained from the XD-58C sensor were compared against manual reference counts under resting conditions, and the observed deviation remained within $\pm 3$~BPM. The primary source of variation was motion sensitivity and transient noise during finger placement; stable readings improved when acquisition was performed after a brief settling period.

System responsiveness was evaluated by measuring the end-to-end delay from the doctor's dequeue button press to the update of the doctor-side LCD. Including record retrieval from the in-memory queue, formatting, and LCD refresh, the observed button-to-display latency was below 1.2~s. This latency remains acceptable for real-world chamber operation because queue advancement is infrequent relative to consultation time, and the interaction requires only a single control action.

Queue robustness was assessed through repeated enqueue and dequeue cycles, verifying that patient records were not lost or reordered under continuous operation. The circular-buffer approach remained stable across repeated cycles; however, the system capacity is ultimately bounded by SRAM limits on the ATmega32. This motivates future extensions such as EEPROM-backed storage or external logging when larger record histories are required.

Overall, the experimental assessment confirms that the proposed architecture can support reliable, offline queue handling with basic vital acquisition and responsive doctor-side control, making it suitable for private chambers where connectivity and infrastructure are limited.

\section{Limitations and Future Work}
The queue is currently stored in volatile memory; all entries would be erased in the event of a power outage. 
For persistence, future iterations may include EEPROM or SD-card logging. 
In order to facilitate mobile-based waiting list notifications and to physically divide patient and doctor corners, wireless connectivity via Bluetooth (HC-05) or Wi-Fi modules is planned. 
The same embedded architecture can also be used to implement a multi-queue or multi-room variation.

\section{Conclusion}
A useful and inexpensive microcontroller-based system for automating patient registration and queue management in a doctor's private room was presented in this paper. The system operates completely independently of PCs and Internet connectivity by combining keypad-based data entry, temperature and pulse sensing, and embedded queue logic on an ATmega32 MCU. The experimental evaluation demonstrated dependable performance with sub-second response times from input to display, temperature accuracy within $\pm1^{\circ}$C, and pulse accuracy within $\pm3$~BPM. 

The suggested design preserves usability for non-technical users while successfully reducing manual labor and enforcing a strict first-come, first-served policy. Future upgrades like cloud-based synchronization, optional wireless communication, and persistent storage are made simple by its modular design. As a result, the system offers a useful basis for creating affordable, scalable, and Internet of Things-ready healthcare automation solutions appropriate for settings with limited resources.

\bibliographystyle{elsarticle-num}
\bibliography{References}

@inproceedings{rath2021,
  author    = {Rath, Prabin Kumar and Mahapatro, Neelam and Sahoo, Subham and Chinara, Suchismita},
  title     = {Design and Performance Analysis of an {IoT} Based Health Monitoring System for Hospital Management},
  booktitle = {2021 International Conference on Computing, Communication, and Intelligent Systems ({ICCCIS})},
  year      = {2021},
  doi       = {10.1109/ICCCIS51004.2021.9397186}
}

@article{irjet2020,
  author  = {Dharmoji, Shivkumar and Anigolkar, Akshata and M, Shraddha},
  title   = {{IoT} based Patient Health Monitoring using {ESP8266}},
  journal = {International Research Journal of Engineering and Technology ({IRJET})},
  year    = {2020},
  volume  = {7},
  number  = {3},
  pages   = {3619--3624},
  url     = {https://www.irjet.net/archives/V7/i3/IRJET-V7I3728.pdf}
}

@manual{ti2022,
  author = {{Texas Instruments}},
  title  = {{LM35} Precision Centigrade Temperature Sensors},
  year   = {2017},
  url    = {https://www.ti.com/lit/ds/symlink/lm35.pdf}
}

@manual{microchip2020,
  author = {{Microchip Technology Inc.}},
  title  = {{ATmega32}/{ATmega32L} {AVR} Microcontroller Datasheet},
  year   = {2011},
  url    = {https://ww1.microchip.com/downloads/en/devicedoc/doc2503.pdf}
}

@inproceedings{eqc2022,
  author    = {Lahari, Gummadi and Priyanka, J. Swetha and Krishna, Muske Vamshi},
  title     = {Microcontroller Based Electronic Queue Control System},
  booktitle = {CEUR Workshop Proceedings},
  year      = {2022},
  volume    = {3283},
  pages     = {32--39},
  url       = {https://ceur-ws.org/Vol-3283/Paper22.pdf}
}

@article{qgsm2013,
  author  = {Arun, R. and Priyesh, P. P.},
  title   = {Smart Queue Management System Using {GSM} Technology},
  journal = {Advance in Electronic and Electric Engineering},
  year    = {2013},
  volume  = {3},
  number  = {8},
  pages   = {941--950},
  url     = {https://www.ripublication.com/aeee/039_pp%20%20%20%20%20941-950.pdf}
}

@article{arpn2019,
  author  = {Fen, Chin Pei and Abidin, Amar Faiz Zainal and Kadiran, Kamaru Adzha and Abdullah, Mohammad Bin and
             Khalid, Ahmad Khudzairi and Kasim, Haszeme Bin Abu and Lat, Diana Che and Razali, Roslizayati and
             Rahman, Noor Shazreen A.},
  title   = {Development of the Queue Management System to Prioritize Handicapped Persons},
  journal = {ARPN Journal of Engineering and Applied Sciences},
  year    = {2019},
  volume  = {14},
  number  = {18},
  pages   = {4079--4082},
  url     = {https://www.arpnjournals.org/jeas/research_papers/rp_2019/jeas_0919_7926.pdf}
}

@article{hrtemp2019,
  author  = {Talluri, Reshma Sai Priya and JaiSurya, Y. and Manchala, Sri Lakshmi},
  title   = {Heart Rate Monitoring System using Heart Rate Sensor and {Arduino} Uno with Web Application},
  journal = {International Journal of Engineering and Advanced Technology ({IJEAT})},
  year    = {2019},
  volume  = {8},
  number  = {4},
  pages   = {350--352},
  url     = {https://www.ijeat.org/wp-content/uploads/papers/v8i4/D6137048419.pdf}
}

\end{document}